\documentclass[sigconf]{acmart}

\AtBeginDocument{%
  \providecommand\BibTeX{{%
    \normalfont B\kern-0.5em{\scshape i\kern-0.25em b}\kern-0.8em\TeX}}}

\setcopyright{acmcopyright}

\acmConference[DeepNLP’20]{}{July 30, 2020}{Xi'an, China}
\acmBooktitle{SIGIR 2020 Workshop on Deep Natural Language Processing for Search and Recommendation (DeepNLP’20), Xi'an, China}
\acmPrice{15.00}
\acmISBN{978-1-4503-XXXX-X/18/06}

\begin{document}

\title{Tag-Aware Document Representation for Research Paper Recommendation}

\author{Hebatallah A. Mohamed}
\affiliation{%
  \institution{Free University of Bozen-Bolzano}
  \city{Bolzano}
  \state{Italy}
}
\email{hemohamed@unibz.it}

\author{Giuseppe Sansonetti}
\affiliation{%
  \institution{Roma Tre University}
  \city{Rome}
  \country{Italy}}
\email{gsansone@dia.uniroma3.it}  
  
\author{Alessandro Micarelli}
\affiliation{%
  \institution{Roma Tre University}
  \city{Rome}
  \country{Italy}}
\email{micarel@dia.uniroma3.it} 

\begin{abstract}

Finding online research papers relevant to one's interests is very challenging due to the increasing number of publications. Therefore, personalized research paper recommendation has become a significant and timely research topic. Collaborative filtering is a successful recommendation approach, which exploits the ratings given to items by users as a source of information for learning to make accurate recommendations. However, the ratings are often very sparse as in the research paper domain, due to the huge number of publications growing every year. Therefore, more attention has been drawn to hybrid methods that consider both ratings and content information. Nevertheless, most of the hybrid recommendation approaches that are based on text embedding have utilized bag-of-words techniques, which ignore word order and semantic meaning. 
In this paper, we propose a hybrid approach that leverages deep semantic representation of research papers based on social tags assigned by users. The experimental evaluation is performed on CiteULike, a real and publicly available dataset. The obtained findings show that the proposed model is effective in recommending research papers even when the rating data is very sparse.

\end{abstract}




\maketitle

\section{Introduction}
\label{sec:int}

With the advent of the Web, researchers worldwide can access the ever-growing archives of research papers. 
On the one hand, this represents an invaluable opportunity for them, on the other hand, however, it makes it increasingly difficult to find articles of their actual interest.
One way that researchers have to find articles is by following citations in other articles that they are interested in, but this way limits the researchers to specific citation communities. Therefore, recommender systems (RSs) are becoming crucial to make effective use of available information. RSs provide an effective solution to the \textit{information overload problem} by automatically capturing user preferences and recommending related information of potential interest to the target user~\cite{Ricci15}. 

In the context of research papers, RSs can allow researchers to organize, find, and even share scholarly publications. There are two ways in which RSs can capture user preferences: explicitly, by enabling the user to enter her preferences, or implicitly, by monitoring her activities such as browsing the Web or reading documents.
RSs are usually categorized into the following categories: \textit{collaborative filtering} systems that attempt to identify groups of people with similar tastes to those of the user and recommend items that they have liked, and \textit{content-based} systems which take advantage of content information to recommend items similar to those previously preferred by the user. Generally, collaborative systems report better performance than content-based approaches, but their success relies on the presence of a sufficient number of user ratings~\cite{adomavicius2005toward}. Research paper RSs suffer from the cold-start problem: most users normally give a limited number of ratings, thus resulting in a sparse user-item rating matrix and making it difficult to summarize users' preferences.

Matrix factorization (MF) is a collaborative filtering based technique, which has become a dominant solution for personalized recommendation~\cite{koren2009matrix}.
It allows RSs to incorporate additional sources of information about items such as item textual representation as for research papers. Since research paper content is large, a good representation based on text is essential. A common practice is to model text in a document as a set of word features~\cite{hassan2017personalized,Heba19}, namely, bag-of-words (BoW). Often, traditional natural language processing (NLP) techniques are applied, such as stop-words removal or stemming, to only keep meaningful features. However, these representation methods do not have the capacity to modeling the semantics embedded in text data. A word can express different meanings and different words can be used to describe the same meaning. Such word ambiguities are often referred to as the \textit{polysemy problem} and the \textit{synonymy problem}, respectively, and the overall phenomenon is known as \textit{vocabulary mismatch problem}~\cite{Furnas87}.

Deep learning models have recently shown great potential for learning effective semantic representation and achieved state-of-the-art performance in many NLP applications. 
In this work, we adopt a model proposed in~\cite{hassan2018semantic} for tag prediction in research papers. The model relies on a hierarchical attention network (HAN)~\cite{Yang16} to encode the semantic representation of the research paper title and abstract for the tag prediction task. Although this approach was proposed for tag prediction, we exploit it for capturing semantic feature representation of papers through tags. After that, we incorporate those tag-aware document representations into an MF framework to derive document rankings for users. We show how semantic document representations based on social tags can be combined with the traditional collaborative filtering methods to yield superior performance with any number of ratings.  
More specifically, the contributions of this article are as follows:
\begin{itemize}
    \item To describe a tag prediction model based on a hierarchical attention network to derive rich semantic representations of scholarly publications;
    \item To propose a novel research paper recommendation approach based on the incorporation of the inferred tag-aware item representations into a matrix factorization framework;
    \item To illustrate a comparative experimental analysis on a real publicly available dataset between the proposed system and state-of-the-art systems.
\end{itemize}

The rest of this paper is structured as follows. Section~\ref{sec:rel} presents an overview of some works related to our research activities. The proposed deep tag-aware matrix factorization approach is illustrated in Section~\ref{sec:met}. In Section~\ref{sec:exp}, we describe the dataset, the evaluation methodology employed to assess the system performance, and the experimental results. Finally, we draw our conclusions and discuss some possible future works in Section~\ref{sec:con}.


\begin{figure*}[hbt!]
\begin{center}
\includegraphics[scale=0.65]{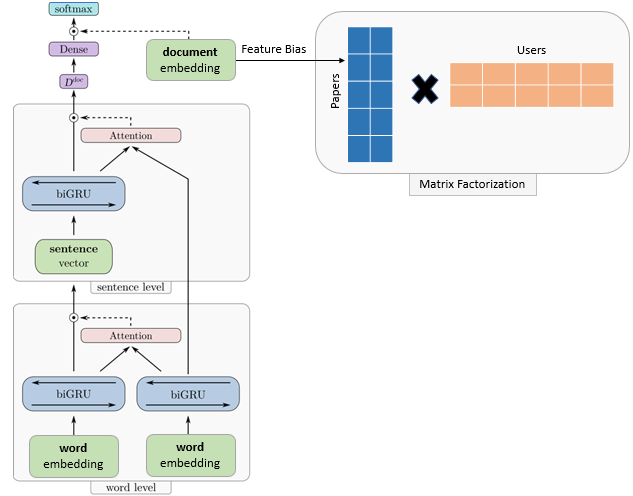}
\caption{Overview of the proposed approach.}
\label{fig:pmf}
\end{center}
\end{figure*}

\section{Related Work}
\label{sec:rel}

To improve the performance of research paper RSs, there have been successful efforts to combine collaborative filtering with content-based filtering approaches.
Pera \textit{et al.} proposed \textit{PubRec}, which makes recommendations based on paper similarities and author ratings~\cite{Pera11}. The paper similarity relies on a word-correlation matrix~\cite{koberstein2006using} to determine the similarity between any pair of tags assigned to their respective papers, in order to capture and represent their contents.
Wang \textit{et al.} proposed the collaborative topic regression (CTR) model that combines a latent dirichlet allocation (LDA) topic model with a probabilistic matrix factorization (PMF)~\cite{mnih2008probabilistic} model for better recommendation~\cite{wang2011collaborative}. Wang \textit{et al.} proposed collaborative deep learning (CDL) that integrates stacked denoising autoencoder (SDAE)~\cite{vincent2010stacked} into PMF.
However, LDA and SDAE do not fully capture document information, as they are based on the bag-of-words model that ignores contextual information of documents such as surrounding words and word orders. Furthermore, they require rich word content which may not be available in research paper recommendation. The full-text of papers cannot be accessed in open-source datasets, and only a relatively short abstract exists in most cases.

Few works have taken advantage of deep neural networks for research paper representation in RSs. Bansal \textit{et al.} proposed a method leveraging recurrent neural networks (RNNs) to encode the text sequence into a latent vector, specifically gated recurrent units (GRUs) trained end-to-end on the collaborative filtering task~\cite{bansal2016ask}. Performance is further improved through multi-task learning, where the text encoder network is trained for a combination of content recommendation and item tag prediction.

Some other works have considered learning semantic document representation through tags. Weston \textit{et al.} proposed a convolutional neural network that learns feature representations for short textual posts using hashtags as a supervised signal~\cite{weston2014tagspace}. Recently, Xiao \textit{et al.} proposed an approach that
takes advantage of label semantic information to determine the semantic connection between labels and documents for constructing label-specific document representation, where a self-attention mechanism is adopted to identify the label-specific document representation from document content information~\cite{Xiao19}. However, this work has not been exploited in the context of RSs.
To the best of our knowledge, the idea of integrating tag-aware document representation based on HAN with a collaborative filtering technique for research paper recommendation has not been addressed before.

\section{Methodology}
\label{sec:met}

\subsection{Problem Definition}

This paper focuses on the task of recommending items associated with the text. The $j$-th text item is a sequence of ${n}_{j}$ words, ${X}_{j}$ = (${w}_{1_j}$, ${w}_{2_j}$, ..., ${w}_{n_j}$). Additionally, the text items are associated with multiple ${T}$ tags provided by users. There are ${N}$ users who have liked some of the text items. The rating provided by user $i$ on item $j$ is denoted by ${r}_{ij}$.  We consider the implicit feedback setting, where we only observe whether a person has viewed or liked an item. The recommendation problem is to find for each user $i$ a personalized ranking of all unrated items.

\subsection{Proposed Approach}

To mitigate the cold start problem in traditional MF, a widely adopted solution is to incorporate additional sources of information about items or users to achieve additional information-based MF.
While tag annotation may be unreliable and incomplete as input features or item metadata, encouraging item representations to be predictive of these tags can yield useful regularization for the collaborative filtering problem. 

Our approach starts with a hybrid MF technique~\cite{kula2015metadata}, we extend an MF model by combining item textual semantic representation obtained from training HAN for predicting research paper tags~\cite{hassan2018semantic}. The model adopts hierarchical bi-GRUs with an attention mechanism to capture important patterns and semantic representations of research papers (i.e., titles and abstracts). It pays word-level attention to tag triggers and sentence-level attention to those sentences containing tags. We then use the learned document embedding as metadata (i.e., bias) for an MF model as illustrated in Figure~\ref{fig:pmf}.

In order to recommend items that one user interacted with to another similar user, we exploit the availability of user/item features and characterize items/users by vectors of latent factors inferred from their features. To briefly review, we model each user/item as a sum of the representations of its associated features and learn a $d$-dimensional representation for each feature value. The probability $\widehat{r}_{ij}$ of an interaction between a user $i$ and an item $j$ is modeled as the \textit{sigmoid} $\sigma$ of the dot product of the user vector and the item vector, along with the bias terms associated with the item (a $1$-dimensional document embedding):
\begin{equation}
\widehat{r}_{ij}=\sigma\left({q}_{i} \cdot {p}_{j}+b_{j}\right)
\end{equation}
where ${q}_{i}$ is the latent representation of user $i$, ${p}_{i}$ is the latent representation of item $j$, and ${b}_{j}$ is the bias term of item $j$. 
We train the model by minimizing the weighted approximate-rank pairwise (WARP) loss~\cite{weston2013learning}.


\section{Experimental Evaluation}
\label{sec:exp}

\subsection{Dataset}

CiteULike was an online platform that allowed registered users to create personal reference libraries by saving academic papers of interest to them~\cite{bogers2008recommending}. It was running from November 2004 until March 2019. The \emph{citeulike-a} dataset\footnote{\url{https://github.com/js05212/citeulike-a}} is built upon this service and consists of papers in the user libraries, user provided tags on papers, and titles and abstracts of them. For each user, the articles that were added in her library are considered as preferred articles. The dataset is very sparse: there are ratings in only 0.22\% of its user-item matrix entries. Table~\ref{tab:db} shows the statistics of the dataset. In this work, we used only the most frequent tags, where ${T}$ = 300. 

\begin{table}[hbt!]
  \caption{Statistics of the \emph{citeulike-a} dataset.}
  \begin{center}
  \begin{tabular}{cccc}
    \toprule
    Users & Papers & Tags & User-Paper\\
    \midrule
    5,551 & 16,980 & 46,391 & 204,986 \\
    \bottomrule
  \end{tabular}
  \label{tab:db}
  \end{center}
\end{table}


\subsection{Baselines}

To perform an empirical evaluation of the proposed method (WARP + HAN), we made a comparison with the following baselines:
\begin{itemize}
\item \textbf{BPR}~\cite{rendle2009bpr}. Bayesian Personalized Ranking, a popular method for recommender systems, which uses pairwise \emph{log-sigmoid} loss. 
\item \textbf{WARP}~\cite{weston2013learning}. Weighted Approximate-Rank Pairwise loss ba\-sed MF model that produces state-of-the-art results in Top-$K$ recommendation. 
\item \textbf{WARP + TFIDF}. A hybrid approach that combines the WARP loss based MF model with items TF-IDF representation.
\item \textbf{WARP + Tags}. A hybrid approach that combines WARP loss based MF model with item tags (in terms of one-hot encoding).
\end{itemize}


\subsection{Setup}

We follow~\cite{hassan2018semantic} for setting the hyper-parameters of the tag prediction model. The maximum number of sentences per document is empirically set to 10, and the maximum number of words per sentence is set to 50. In~\cite{hassan2018semantic}, the authors predict the most frequent 10 tags. In our experiments, we train the model for predicting the most frequent 300 tags.
We use the \emph{LightFM}\footnote{\url{https://github.com/lyst/lightfm/}} implementation of BPR and WARP, which is a popular Python library. We use it also for the implementation of the hybrid approaches. Similar to~\cite{wang2011collaborative}, we set the latent factor dimension to 200. The regularization parameters is set to $1e-05$. The maximum number of negative samples for WARP fitting is set to 100. Finally, the number of epochs to train the MF models is selected by searching between 50 and 200, the best performance is achieved by using 100 epochs.


\subsection{Evaluation Metric}

Starting from the \emph{citeulike-a} dataset, we randomly selected $P$ items associated with each user to form the training set and we used all the rest of the dataset as the test set. In our experiments, we set $P$ to 10.
As in~\cite{wang2011collaborative,wang2015collaborative}, we used Recall as the performance measure since the rating information is in the form of implicit feedback. This means a zero-entry could be since the user is not interested in the item, or that the user is not aware of its existence. Hence, Precision is not a suitable performance measure. And like most of RSs, we sort the predicted ratings of the candidate items and recommend the top-$K$ items to the target user. The Recall@$K$ for each user is then defined as follows:
\begin{equation}
\text{Recall}@K=\frac{\text{number of items the user likes among the top-$K$}}{\text{total number of items the user likes}}
\end{equation}
The final result reported the average Recall on all users.


\subsection{Results}
\label{sec:res}

Table~\ref{fig:results2} summarizes the Recall@K for all the models, where ranks $K$ = 50, 100, 150, and 200 are selected. We first note that WARP outperforms BPR. Then, we observe that incorporating TF-IDF as item features increases the Recall significantly. While incorporating tags information also gives better recall in comparison to BPR and WARP, but it is not outperforming the TF-IDF based approach. More importantly, we note that our proposed model, WARP + HAN, outperforms all the four baselines, which shows the effectiveness of integrating deep semantic representation of research papers based on social tags with collaborative filtering.

\begin{table}[hbt!]
  \caption{Performance comparison among the different models in terms of Recall@$K$. The improvement over baselines is statistically significant on a paired $t$-test $(p\ll0.01)$.}
  \label{fig:results2}
  \begin{tabular}{l|cccccc}
    \toprule
    Models & @50 & @100 & @150 & @200 &  \\
    \midrule
    BPR & 0.04 & 0.08 & 0.11 & 0.13  \\
    WARP & 0.07 & 0.11 & 0.14 & 0.16  \\
    WARP + TFIDF & 0.09 & 0.15 & 0.19 & 0.22 \\
    WARP + Tags & 0.09 & 0.13 & 0.16 & 0.19 \\
    \midrule
    \textbf{WARP + HAN} & \textbf{0.10} & \textbf{0.17} & \textbf{0.23} & \textbf{0.27} \\
    \bottomrule
  \end{tabular}
\end{table}


\section{Conclusion}
\label{sec:con}

In this paper, we described a personalized research paper recommendation approach, which integrates the matrix factorization technique with tag-aware document embeddings. The embeddings are extracted from a social tag prediction model that exploits bidirectional gated recurrent units (bi-GRUs) and attention mechanism, for aggregating important words and sentences based on the assigned tags, to increase the general representation and visualization of the key concepts in research papers. The results of our experiments on the CiteULike dataset show that the proposed approach outperforms state-of-the-art collaborative filtering based techniques. A limitation of our research is that we only considered the most frequent 300 tags for generating the document representations. In the future, we would like to perform the experiments with a larger number of tags. Finally, we would like to further compare our proposed approach with multi-task learning, where we train one model for both the tag prediction and paper recommendation tasks.

\bibliographystyle{spmpsci}     
\bibliography{main}

\end{document}